# Switchable Topological Phase Transition and Novel Nonlinear Optical Properties in ReC$_2$H Monolayer

Chunmei Zhang[1#*], Hanqi Pi[2#], Liqin Zhou[2#], Si Li[1], Jian Zhou[3], Aijun Du[4], Hongming Weng[2,5]*

[1]School of Physics, Northwest University, Xi'an 710069, China

[2]Beijing National Laboratory for Condensed Matter Physics and Institute of Physics, Chinese Academy of Sciences, Beijing 100190, China

[3]Center for Alloy Innovation and Design, State Key Laboratory for Mechanical Behavior of Materials, Xi'an Jiaotong University, Xi'an, China.

[4] School of Chemistry and Physics, Queensland University of Technology, Gardens Point Campus, Brisbane, QLD 4001, Australia

[5]Songshan Lake Materials Laboratory, Dongguan, Guangdong 523808, China

Corresponding author: chunmeizhang@nwu.edu.cn
hmweng@iphy.ac.cn

C. Z, H.P, and L. Z. contribute equally to this work.

## Abstract

Extensive investigations on topological phase transition (TPT) in three-dimensional compounds have been done. whereas, rare in two-dimensional systems, let alone noncentrosymmetric materials. In this work, based on first-principles calculations, we explore an inversion symmetry broken structural ReC$_2$H monolayer. We reveal that it undergoes two TPTs, namely from normal insulator to Z$_2$ topological insulator and back to normal insulator, at the critical biaxial strain of 2.3% and 7.8%, respectively. The band inversion occurs at the generic momentum in the first TPT, while at high symmetric K point in the second one. Usually, band inversion is identified by the exchange in the components or irreducible representations of the wavefunctions. These quantities can be easily obtained in theoretical calculation but hard to be detected in experimental techniques like Angle-resolved photoemission spectroscopy. It is well known that nonlinear optical (NLO) response is very sensitive to the components and symmetries of the engaged bands, which also incorporates information of band topology. Therefore, we study the shift current, one of the widely explored NLO responses in noncentrosymmetric systems, during the two TPTs. We find that in both cases band inversion leads to the sign change of shift vectors around the momenta where the bandgap closes and reopens. Whereas the shift current, as the overall contribution of shift vectors weighted by the absorption rate in the whole Brillouin zone, may keep its direction. This work offers insight that a scrutinized examination is highly demanded in utilizing shift current to detect TPT.

**Introduction**

Topological materials featuring topological invariants have been classified as topological insulator (TI), topological crystalline insulator and topological semimetal (TSM). Among them, the TI has $Z_2$ invariant protected by time-reversal symmetry and the corresponding gapless surface Dirac state. TSMs are characterized by having the conduction band crossing with the valence band at Fermi level and the crossing points can be identified by topological charge as the invariant. The typical TSMs include Dirac semimetal and Weyl semimetal (WSM), which can be viewed as the critical phase during the topological phase transition (TPT) between TI and normal insulator as proposed by S. Murakami.[1] The TPT can be driven by external stimulus like electric field[2], magnetization[3], strain and pressure[4,5]. Essentially, these methods control the mass parameter '$m$'[1] to tune the band inversion accompanied by the Dirac semimetal or WSM in system with or without inversion symmetry. Thus, during TPT the system usually exhibits various exotic physical phenomena, which can be used as a guide for identifying topological materials in practice. For example, the insulator-metal transition and enhanced diamagnetism when temperature drops down have been observed in $ZrTe_5$ and $HfTe_5$.[6] Therefore, exploring more unique and intriguing phenomena related with TPT is of great significance. Since linear and nonlinear optical (NLO) properties are engaged with two or more bands and they are directly determined by the symmetries and components of their wavefunctions, it is believed that the band inversion will likely be traced by inspecting the optical responses during TPT.[7,8,9] Another advantage of utilizing optical approaches is its noncontacting and highly-tunable nature, which does not require mechanical or electrochemical methods to directly contact the systems during operation. Hence, it is less susceptible to lattice damage and does not introduce unwanted impurities or disorders during material fabrication processes. The tuning of light frequency, incident angle, polarization, and intensity can be easily performed without directly affecting the samples.

One of the NLO responses is bulk photovoltaic effect (BPVE)[10,11,12], which yields steady-state electric current under homogeneous illumination upon a single noncentrosymmetric crystal,[13] and the fabrication of heterostructure, such as p-n junction, to separate the electron and hole photocarriers is avoided. It offers great potential in energy harvesting, photodetection, rectification, etc.[14] Above bandgap photon irradiation generates electronic interband transition[7], which is governed by geometry features of wavefunctions such as Berry connection and quantum metric dipoles[15,16,17] in momentum space. The primary BPVE is the generation of shift current under linearly polarized light in time-reversal symmetry systems, which arises from the displacement of carrier positions, namely the center of Wannier functions, during interband optical excitation[18]. Thus, the topological nature of shift current can be expressed via Berry connection, which is looked as the expectation value of the position operator in the crystal unit cell. Described by the formula composed by Berry connections of the two bands involved in the resonance, the shift vector is a gauge invariant quantity measuring the difference of intracell coordinates between the two bands, which is odd and flips sign under band inversion referring to the interchange of the conduction and valence band index.[18, 19] As the net shift current is the product of absorption rate and shift vector, one may expect that the change of photocurrent magnitude and/or direction could be utilized to detect the band inversion during TPT.[7]

Generally, TIs possess intrinsic band inversion and enhanced Berry connection near their bandgap,[20, 21, 22, 23, 24] which could significantly boost their optical responses, linear[24] or nonlinear.[7, 9] In addition, WSM happens to harbor inversion asymmetry structure, and the low energy joint density of states (JDOS) scales as $\omega^2$ ($\omega$ being the frequency of the incident light).[25] Due to the singular Berry connection in WSMs, the shift-current response in the low energy limit is anomalously large.[25, 26] Similar designing principle for searching shift current materials points to specific systems with semi-Dirac type of Hamiltonians and strong singularity in JDOS.[27] To the materials realization perspective, experimentally, a large shift current is detected (154±17 μA/V$^2$ at the photon energy $\hbar\omega$ = 0.117 eV) in TaAs[25] due to the gapless band structure and anomalous Berry connection in WSM. Besides, Q. Xu, *et al.*[23] have theoretically discovered 46 WSM candidate materials based on high-throughput calculations and most of them exhibit giant shift current feature.

The above topological view offers a route to design large NLO materials. However, there are still several problems remain unsolved: (i) Most investigations are on three-dimensional (3D) WSMs, considering that two-dimensional (2D) materials often show stronger van Hove singularities in the density of states at the band edge,[28] much enhanced shift current is expected. Unfortunately, 2D noncentrosymmetric TI with controlled TPT is poorly studied either in the point view of materials design or for NLO effect study; (ii) Since shift current is the overall contribution of shift vectors weighted by the absorption rate in the whole Brillouin zone (BZ), the reversal of shift vector direction at some momenta could not guarantee the direction change of shift current, especially when the band inversion momenta barely contribute to it. Nevertheless, these have not been carefully discussed.

In this work, we address these problems based on density functional theory (DFT) calculations,

by investigating a novel 2D system, single-side hydrogenated rhenium carbide (ReC$_2$H) monolayer with $C_{3v}$ symmetry. The inversion asymmetry together with the strong spin-orbit coupling (SOC) generates giant Rashba spin splitting at generic $k$ points. The SOC bandgap is direct, controllable, and continuously tuned by external stimuli. We show that the system experiences two TPTs under intermediate biaxial strains: from a normal insulator to a TI at a critical strain ($\varepsilon_{c1}$) around 2.3%, and back to a trivial insulator when strain is over 7.8% ($\varepsilon_{c2}$). During the two TPTs, the dependence of shift current is also analyzed. The shift current direction is unchanged across $\varepsilon_{c1}$, while it reverses at $\varepsilon_{c2}$. This has been analyzed and it is noted that the band inversion can cause reverse of shift vectors locally in the momentum space while the shift current is determined by integration weighted by absorption rate over the whole space.

## Results

**Geometric structure of ReC$_2$H monolayer**

The experimental crystal structure of ReC$_2$[29] is shown in Fig. 1a. It crystalizes in hexagonal structure with space group $\bar{P}6m2$ (No. 187), and presents trivial metal characters.[29] When it is H-passivated on a single side, a single layer ReC$_2$H (Fig. 1b-c) belongs to space group $P3m1$ (No. 156). The optimized lattice constant is 3.002 Å. The ReC$_2$H monolayer resembles MoS$_2$ monolayer which lacks inversion symmetry. A single side H passivation further breaks the out-of-plane mirror symmetry.

From a material realization perspective, H passivation engineered surface have been fabricated and studied extensively[30], i.e., atomic-scale patterning of H terminated Si[31, 32] and Ge[33] by scanning tunneling microscopy or single layer H-functionalized germanane (GeH monolayer) grown onto SiO$_2$ substrate.[34] In our system, H atoms passivate the C dangling bonds, which not only act as a resist in analogy to that used in optical lithography[33] but also tune the chemical potential. The H adsorption energy is -1.094 eV/atom, indicating the strong interaction (chemical bonding) between the ReC$_2$ and H atom. Also, the crystal structure of ReC$_2$H monolayer resembles that of the 2H phase MoS$_2$ which has been synthesized by chemical vapor deposition long before.[35, 36] In addition, Figure s1a presents the calculated phonon spectrum for the ReC$_2$H monolayer, where no imaginary frequency can be seen in the BZ, confirming its dynamic stability. The *ab initio* molecular dynamics (AIMD) simulations are further carried out for 10 ps at 300 and 500 K (as shown in Figure s1b, c), and no evidence of structural destruction suggests thermal stability of monolayer ReC$_2$H. Thus, there is a great synthesis possibility in experiments for ReC$_2$H monolayer.

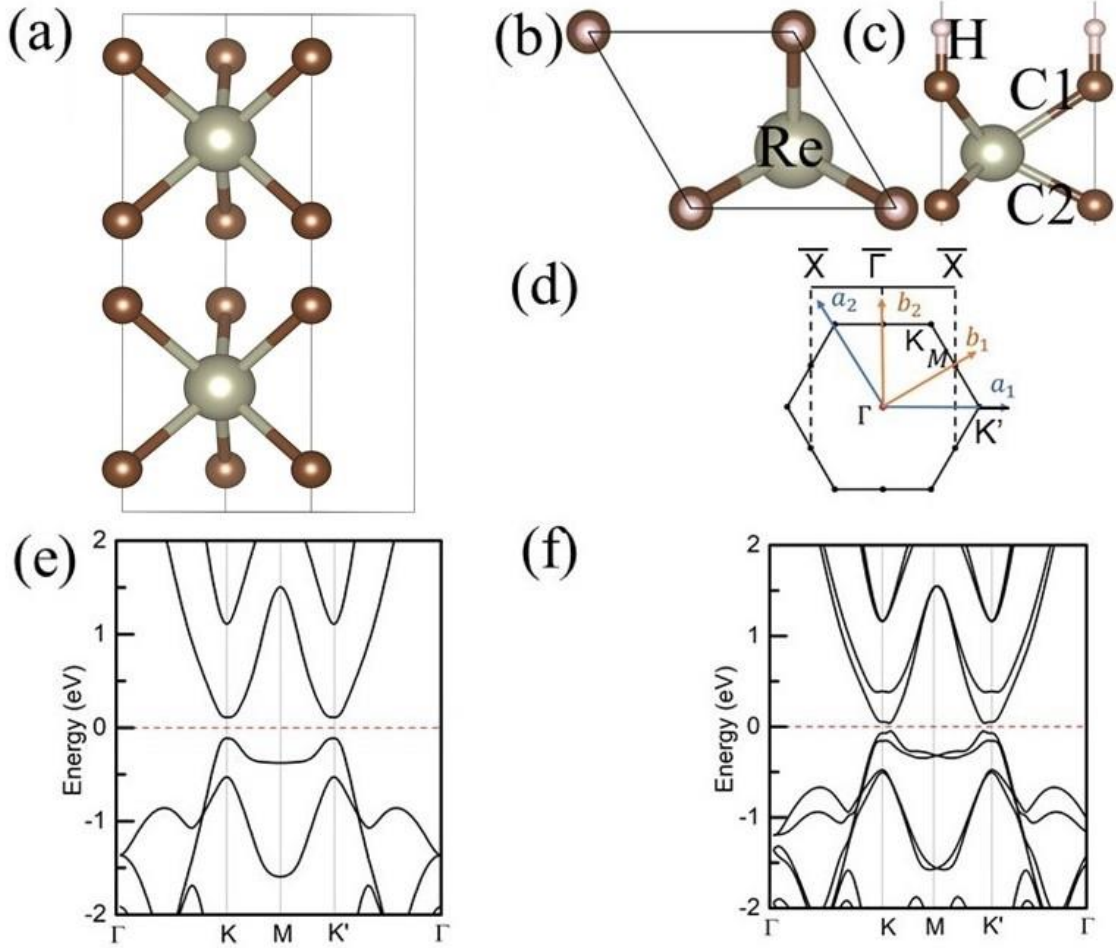

Figure 1. (a) Side view of 1x1x2 supercell ReC$_2$ compound, (b-c) top and side views of ReC$_2$H monolayer, where C1 and C2 are labelled to differentiate C atoms. (d) The 2D BZ and the projected one-dimensional BZ for armchair edge. The band structure of ReC$_2$H monolayer calculated at the GGA level of theory (e) without SOC and (f) that with SOC. The Fermi level is denoted by red dashed lines.

**Electronic structure analysis.**

We first calculate the band structure of ReC$_2$H without SOC by GGA and plot it along the high-symmetry k-path in Fig. 1e, which indicates that it is a semiconductor with bandgap of about 0.3 eV. The bands near the Fermi level are mainly contributed by C2 (unsaturated carbon) $p_z$ orbital and Re $d_{xy}$, $d_{x^2-y^2}$ orbitals. Including SOC would break the spin rotational symmetry and trigger the Rashba like spin splitting at generic $k$ point[37] (Fig. 1f). Note that the spin degeneracy at Γ and M results from the Kramer degeneracy at time reversal invariant momenta. The Rashba splitting reduces the bandgap to ~0.1 eV. This could reduce the potential critical field to induce TPT. Thus, we mainly focus on the discussion of the electronic structure with SOC in the following.

When the space inversion symmetry is broken, the topological invariant $Z_2$ for a band insulator with time-reversal symmetry can be obtained through the evolution of Wannier charge centers (WCCs) of occupied states as shown in Fig. s4[38, 39]. The $Z_2$ invariants under different strain are shown in Fig. 2, together with part of the valence and conduction bands along $\Gamma - K - M$ around $K$ point. The whole band structures along high symmetry line $\Gamma - K - M - K' - \Gamma$ are given in Fig. s3. The bands near Fermi level are mainly contributed by Re $d_{xy}$, $d_{x^2-y^2}$ orbitals (weighted by green curve) and C2 $p_z$ orbital (weighted by red curve) in Fig. 2a-l. We find the bandgap closes and reopens twice, indicating the first TPT is around the critical strain of $\varepsilon_{c1}$ = 2.3% and the second one is around $\varepsilon_{c2}$ = 7.8%. Detailed analysis reveals that the first TPT occurs at generic moment around the path K-M, and the second one is at $K$ point. Before the second TPT with strain $\varepsilon_{c2}$, the irreducible representations (IR) for VBM and CBM are GM5 and GM6, respectively. The energy order of them is inversed at $\varepsilon_{c2}$. More details for the IR of bands at K point are listed in Table s1. It should be noted that there is also band exchange in IRs around the strain of 7% (Fig. 2i) in the two conduction bands. Although it has no effect on the topological property of the occupied states, it enables the second TPT.

The band inversion identified by IRs of related bands has been seen in the second TPT, while this method fails in analyzing the first TPT since the CBM and VBM at generic moment have the same IRs. On the other hand, the orbital components of them resemble each other as shown in Table s3. Therefore, the first TPT is not easy to be captured in above two methods although $Z_2$ changes from 0 to 1. The strain induced TPT of ReC$_2$H can be understood with the theory proposed by Murakami, *et al.*[1] Around the gap-closing point, the $\boldsymbol{k} \cdot \boldsymbol{p}$ Hamiltonian reduces to a two-component Hamiltonian $H = m\sigma_z + (k_x - k_{0x})\sigma_x + (k_y - k_{0y})\sigma_y$. Taking the first TPT as an example, the mass parameter $m$ changes sign around the strain of 2.3% accompanied by a TPT. The TPT point is at ($k_{0x}$, $k_{0y}$) when $m$ equals to 0. In general, the location of TPT point can be at any generic momentum. According to the space group $P3m1$ of this system, in the first TPT there are all together 12 gap-closing points that are related by C$_3$ rotation, mirror symmetry and time-reversal symmetry. In the second TPT, the gap-closing points are at K and K'.

As the gapless topological boundary state is an essential character of a TI belonging to $Z_2$ classification, we further calculate the one-dimensional (1D) edge states. Both armchair (Fig. 2m-o) and zigzag (Fig. s5) edge states are plotted for the ε = 0.9%, 4%, and 10%, respectively. The 1D Dirac cones of the edge state for 0.9% and 10% strained ReC$_2$H monolayers connect valence band only, indicating the system trivial with $Z_2=0$. On the contrary, two helical edge states protected by time-reversal symmetry are visible clearly at the strain of 4%, where the Dirac cone at time-reversal invariant momentum $\bar{X}$ are demanded by Kramer degeneracy and its two branches connecting the projections of bulk valence and conduction bands, respectively.

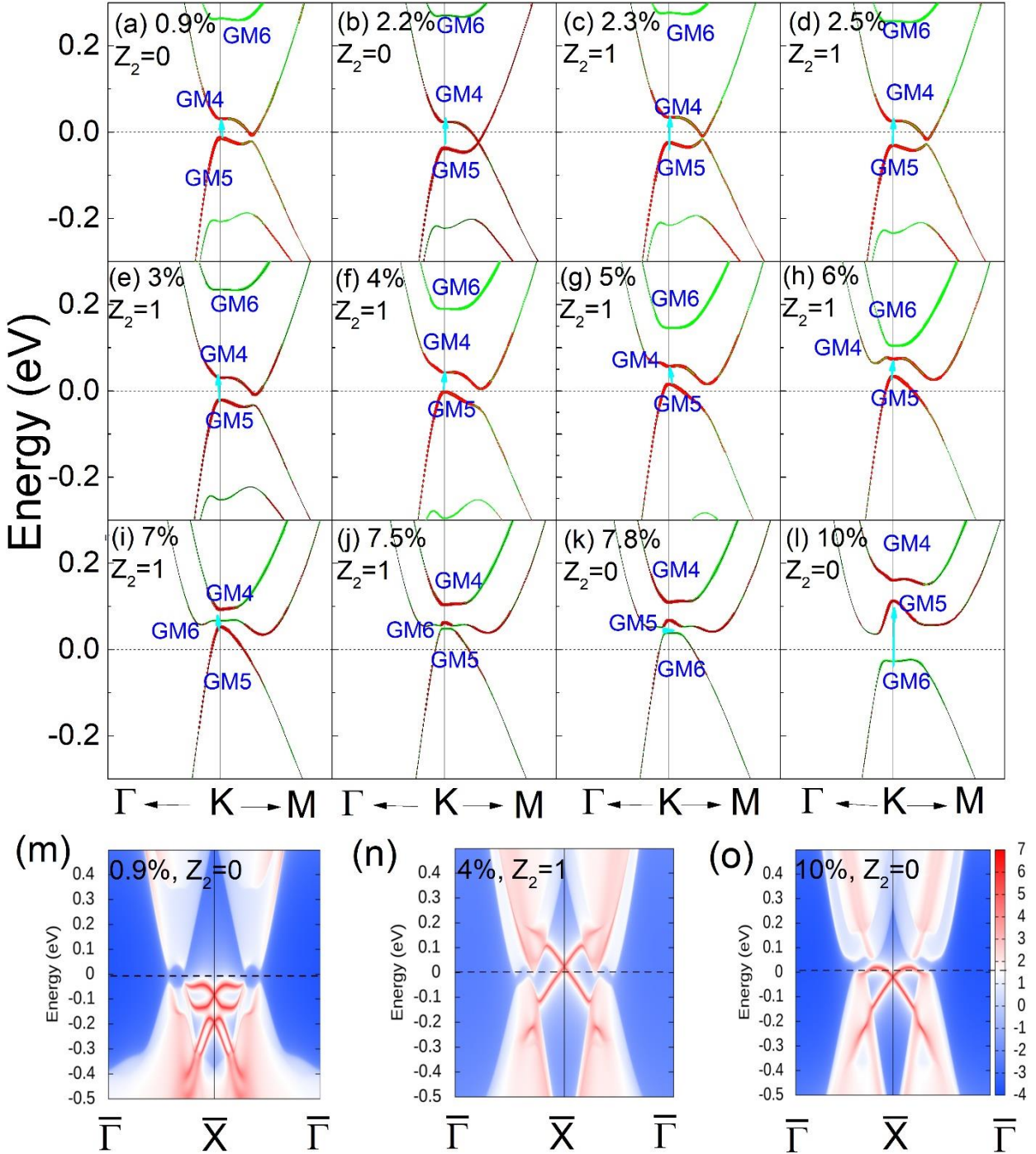

Figure 2 (a-l) Band structures and $Z_2$ invariant for ReC$_2$H monolayer under 0.9%-10% strain. The red curve represents the weight of C2 $p_z$ orbitals in the wavefunction of the bands, and the green curve denotes the weight from Re $d_{xy}$, $d_{x2-y2}$ orbitals. The band structure is calculated along high symmetry path $\Gamma - K - M$, and only that centering K point is shown. (m-o) 1D armchair edge states of 0.9%, 4%, and 10% strained ReC$_2$H monolayers. The Fermi level is marked by black dashed line.

## Shift current variation under topological phase transition

While both theory and experiment demonstrate that topological materials show large NLO response[7,40], we further consider the influence of band topology on the shift current. In our system, two TPTs occur, one with explicit band inversion changes while the other does not. It would be interesting to investigate how the TPTs would influence the NLO.

The inversion symmetry broken monolayer ReC$_2$H has a mirror plane perpendicular to $\hat{x}$, which is along the zigzag direction. The mirror and C$_3$ rotation symmetries lead to four independent non-vanishing shift current response tensors, namely in-plane component: $\sigma^{yyy}=-\sigma^{yxx}=-\sigma^{xxy}$ and out-of-plane ones: $\sigma^{zxx}=\sigma^{zyy}$, $\sigma^{zzz}$, $\sigma^{xxz}=\sigma^{yyz}$. In this 2D system, since the out-of-plane conductivity component is not a well-defined quantity, we mainly focus on the discussion of in-plane component $\sigma^{yyy}$. Around the first and second TPTs, $\sigma^{yyy}$ variation under incident photon energy is calculated and shown in Fig. 3a and b, respectively. It is noticed in both figures that the first peak or valley of $\sigma_{yyy}$ is mainly contributed by the transition between the VBM and CBM around K point, since the corresponding photon energy is comparable with the direct bandgap value at K shown in Table S2 and the cyan arrow in Fig. 2. Across the first TPT, the incident photon energy of the first $\sigma_{yyy}$ peak decreases continuously, while the peak value increases with strain from 0%—2.3% and then decreases after $\varepsilon_{c1}$. The maximum value reaches 600 $\text{Å} \cdot \mu A/V^2$ (at ℏω=0.070 eV and ε = 2.3%). Before and after this TPT, the sign of $\sigma_{yyy}$ keeps unchanged. On the other hand, around the second TPT (6%-12%), $\sigma_{yyy}$ peak value decreases from ~500 $\text{Å} \cdot \mu A/V^2$ (ε = 6.0%) to ~70 $\text{Å} \cdot \mu A/V^2$ (ε = 7.8%). After the TPT, $\sigma_{yyy}$ reverses its sign and gradually increases its value from -150 to ~-200 $\text{Å} \cdot \mu A/V^2$ (at ε = 10%). Before the second TPT, the first peak $\sigma^{yyy}$ value keeps decrease from ε =6% to 7.8%. This counterintuitive shift current value changes should be noted. The reason is that the first peak position of $\sigma_{yyy}$ is not exactly located at K point, which is slightly away from K point.

We further move to analyze the band topological effect on the shift current direction. As described in Method section, the shift current can be expressed as the integral of the product of the absorption rate $|r_{mn}^b|^2 \delta(\omega_{nm} - \omega)$ and the shift vector $R_{mn}^{a,b} = \partial_a \phi_{mn}^b - A_{mm}^a + A_{nn}^a$, where $m$ and $n$ are the band indices. The sign of the absorption rate is always positive, while under band inversion, the shift vector $R_{mn}^{a,b}$ changes to $R_{nm}^{a,b}$ with sign changes. Thus, one expects that TPT prefer to flip the shift current direction. This simple assumption is consistent with calculations in the second TPT in our system (Fig. 3b). However, it is oversimplified at the first TPT, where the shift current direction is not changed as shown in Fig. 3a.

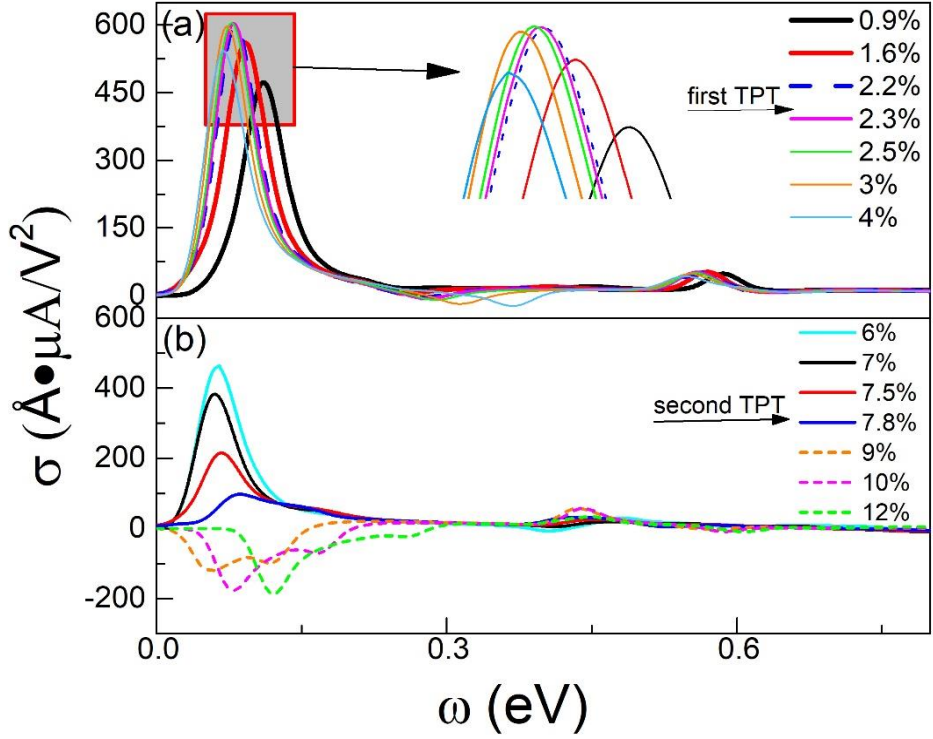

Figure 3. Shift current conductivity $\sigma_{yyy}$ around the first (a) and second (b) TPTs changes with the incident photon energy under different strain.

The distinct shift current direction behavior across two TPTs are novel, and we elaborate on their possible mechanism below. The $k$-resolved shift vector $R_{cv}(k)$ in the BZ only including the highest valence and the lowest conduction bands are plotted for the ε = 2.2% and 2.3%, as shown in Fig. 4a and b, respectively. The first TPT happens between them. The k points for the gap-closing are marked as 1, 2, and 3. For clarity reason, we plot shift vector curves along $k_y$ passing through these points in Fig. 4c and d. It is clearly seen that the shift vector $R_{cv}(k)$ reverses their sign across the first TPT when strain changes from 2.2% to 2.3%. This indicates that the shift vector $R_{cv}(k)$ sign is locked with band inversion during TPT. We compare the k-resolved shift current for the 0.9% ($Z_2$=0) and 4% ($Z_2$=1) with incident light at the corresponding bandgap (ℏω=0.053 and 0.037 eV in Fig. 4e, f, indicating their band edge contributions) and at K point (ℏω=0.090 and 0.065 eV in Fig. 4g, h). Fig. 4i shows the enlarged shift current distribution around K for Fig. 4e-h. It clearly shows that after TPT, the shift vector $R_{cv}(k)$ indeed flips its sign around TPT point (Fig. 4i-1, 2), but the absorption rate is negligibly small compared with that around K point (Fig. 4h-3, 4) where no band inversion occurs. It coincides with the fact shown in Fig. 3a that the peak keeps moving towards lower frequency after the bandgap reopens, being different from the proposal in Ref. [7]. This is due to the large contribution of K point at which the bandgap continues to decrease towards the second TPT. While one should note that $k$ points in BZ have different contributions[41] to the shift current as the absorption rate $|r_{mn}^b|^2 \delta(\omega_{nm} - \omega)$ is also $k$ dependent. With an intensive band structure study across the first TPT, we find that there is van Hove singularity like valence and conduction band edge at K (Fig. 2a-f), which leads to much larger JDOS than the TPT point (Fig. s5). Thus, the main contribution to shift current is from K, rather than from points around the band inversion for the first TPT.

In comparison, the shift current behavior at the second TPT agree well with the previous conclusion. The band inversion happens at K point where shift vector flips sign (Fig. 4j, k) and the bands around K point also contribute mainly to the shift current (Fig. 4l, m). The photon energy adopted in Fig. 4j-m is the same as the bandgap at K point for 7% and 10% strained ReC$_2$H monolayer.

Finally, we compare the value of the shift current conductivity of ReC$_2$H with previously proposed potential 2D systems: ~150 $Å \cdot \mu A/V^2$ for GeS (at a photon energy of ~2.500 eV)[41, 42], ~700 $Å \cdot \mu A/V^2$ for WS$_2$[41] (at an excitation energy of ~2.719 eV ), 220 $Å \cdot \mu A/V^2$ for Bi (110) monolayer (at a photon energy of 0.670 eV) [43], ~8 $Å \cdot \mu A/V^2$ for 2H MoS$_2$ monolayer (with photon energy of 2.800 eV) [44]. Thus, the shift current conductivity is quite large for ReC$_2$H monolayer.

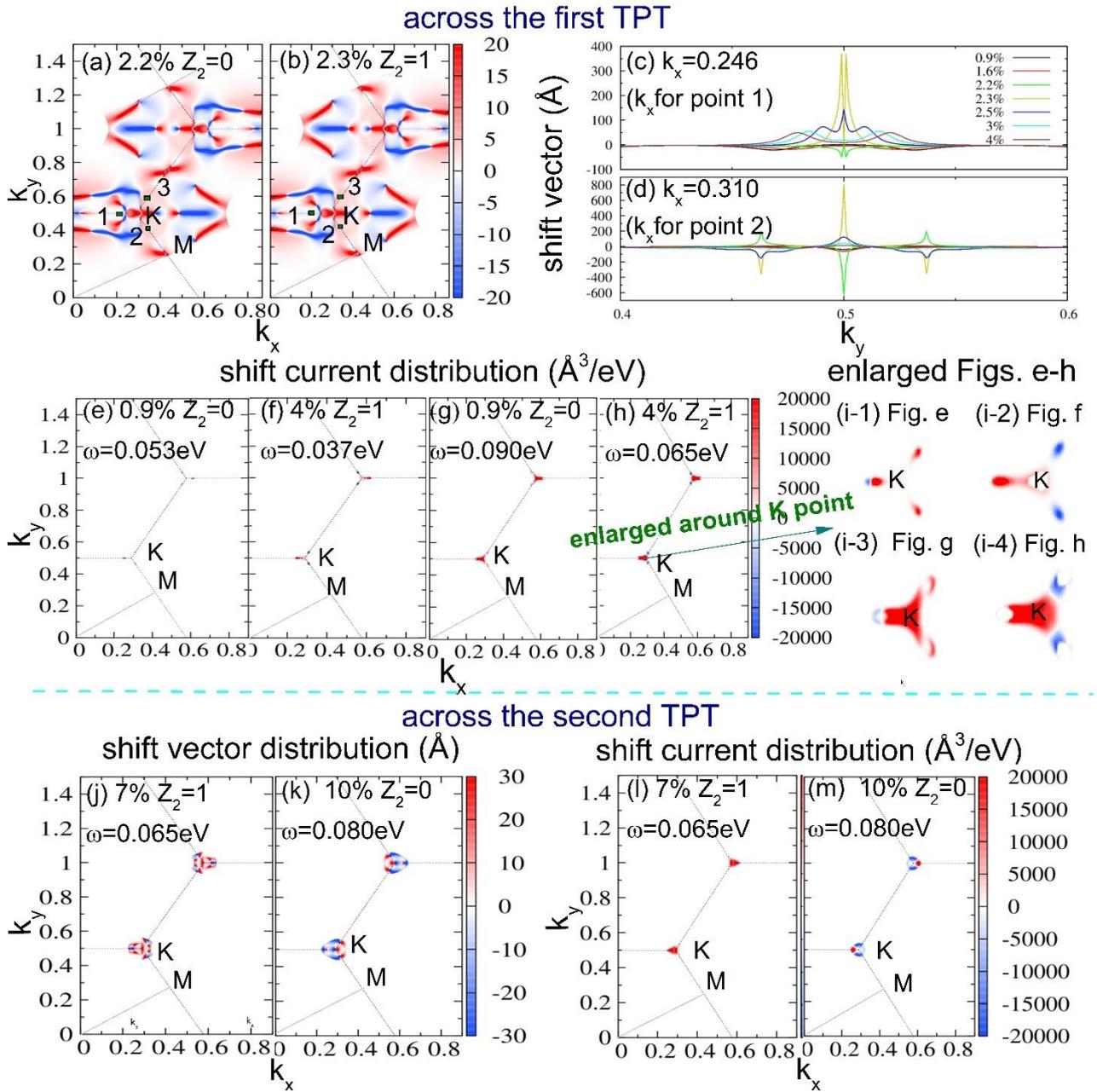

Figure 4 (a-b) The shift vector distribution $R_{cv}(\mathbf{k})$ for ReC$_2$H monolayer with ε = 2.2% and 2.3%, which only includes the contribution from highest valence band and the lowest conduction band. The marked points 1, 2, 3 are along K-M line, which are close to the first TPT point. (c-d) The shift vector distribution along $k_y$ with specified $k_x$ value the same as that of points 1 and 2. (e-h) Distribution of shift current in the whole BZ for 0.9% and 4% strained ReC$_2$H monolayer. The incident light with excitation energy ($E_{nm} = \hbar\omega_{nm}$) of 0.053 eV, 0.037 eV (which is the bandgap near TPT point) and 0.090 eV, 0.065 eV (bandgap at K point) are used for 0.9% and 4% strained ReC$_2$H monolayer, respectively. (i) The enlarged figures of shift current distribution around K for 0.9% and 4% strained ReC$_2$H monolayer. Distribution of (j-k) shift vector and (i-m) shift current in the whole BZ for 7% and 10% strained ReC$_2$H monolayer. The incident light with excitation energy of 0.065 eV and 0.080 eV are used, being the same as the bandgap at K point for the strain of 7% and 10% cases. When we calculate the shift current, the fermi level is shifted to be inside of the bandgap between the highest valence and the lowest conduction band at K point.

## Discussion

In conclusion, we investigate the interplay between TPT and NLO shift current generation in a 2D noncentrosymmetric ReC$_2$H monolayer. Under biaxial strain, the band inversion emerges twice at generic momenta and K points, respectively, which enables a topologically trival-nontrival-trival phase transition with the critical strain of 2.3% and 7.8%. For its NLO response, the phenomenon of enhanced shift current conductivity related to band inversion at both TPTs are witnessed in line with other references. The shift current conductivity reaches 600 $Å \cdot \mu A/V^2$ at an incident photon energy of 0.070 eV, comparable to previous proposed 2D materials. Remarkably, the shift current reverses direction at the second TPT but not the first one as the absorption rate is momenta dependent. Our system with two TPTs offers the first material realization, which evidences that TPT is accompanied by the shift vector sign change but not necessarily flipping the direction of shift current. Thus, a careful analysis is required in utilizing shift current to detect TPT. One the other hand, if k-resolved shift vector could be measured by some experimental technique, it would be a promising way to characterize TPT. Other optical rectification processes are not included in this study, and their effects will be discussed elsewhere. The combination of inversion asymmetric polar topological insulator, giant NLO effect, and the relationship between TPT and shift current could generate considerable impact in the field of optical detector and sensor, energy harvesting and transfer, and terahertz wave generation soon and demonstrate the limitation in using TPT for photocurrent direction control.

## Methods

The DFT calculations are performed with the QUANTUM ESPRESSO package[45], using the projector augmented wave (PAW) potential[46] in PSLIBRARY,[47] the Perdew-Burke-Ernzerhof (PBE) exchange-correlation functional,[48] and a plane-wave basis with energy cutoff of 60 Ry for valence electrons. A vacuum region of about 15 Å along the z direction is adopted to eliminate the artificial layer interactions. The BZ integration is sampled by using Γ-centered Monkhorst–Pack k-point sampling with a grid of 11 × 11 × 1. The van der Waals (vdW) interaction is described by the DFT-D3 method.[49] The WANNIER90 code package[50, 51] is used to construct the tight-binding model based on the maximally localized Wannier functions (MLWFs). The edge states are calculated by using the WANNIERTOOLS software package based on the semi-infinite Green function method.[52] The Wannier interpolation method[53] is adopted to calculate shift current conductivity using a refined $k$ grid of $2000 \times 2000 \times 1$ and the value for the broadening factor of Dirac delta function is η =

0.02eV. The k-point interpolation meshes are tested to obtain the well-converged shift current spectra. The SOC is included self-consistently throughout the calculation. The irreducible representations of electronic bands are calculated in *irvsp* code.[54] In order to explore the dynamic stability, phonon dispersion is obtained by the finite displacement method[55] as implemented in the Phonopy code.[56] The AIMD simulations are carried out with a canonical ensemble at temperature of 300 K and 500K for 10 ps with a time step of 1 fs.[57] Geometry relaxation and electronic structure calculation are also performed using the Vienna ab initio simulation package (VASP), and consistent results are obtained.[58, 59, 60]

In time-reversal symmetry system, under linearly polarized light with electric filed $E$ at frequency $\omega$, the shift current conductivity is evaluated via[12, 61] $\sigma_{bb}^a(\omega) = \frac{\pi e^3}{\hbar^2} \int \frac{d^3k}{8\pi^3} \sum_{m,n} f_{m,n} R_{mn}^{a;b} |r_{mn}^b|^2 \delta(\omega_{nm} - \omega)$, where light is linearly polarized along the *b* direction, and $f_{m,n} = f(E_n) - f(E_m)$ is the difference of Fermi-Dirac occupation. $r_{mn}^b$ is inter-band Berry connections defined as $r_{mn}^b = i\langle m|\partial_{k_a} n\rangle$. $R_{mn}^{ab} = \partial_a \phi_{mn}^b - A_{mm}^a + A_{nn}^a$ is the shift vector and $\phi_{mn}^b$ is the phase of $r_{mn}^b = |r_{mn}^b| e^{i\phi_{mn}^b}$. $R_{mn}^{a,b}$ has the unit of length and can be physically interpreted as position change of a wave packet during its transition from band *m* to band *n*. The $|r_{mn}^b|^2 \delta(\omega_{nm} - \omega)$ evaluates absorption rate from band *m* to band *n* according to the Fermi's golden rule.

## Acknowledgement

We acknowledge the financial support from the National Natural Science Foundation of China (NSFC) under Grants No. 12004306, No. 11974270, No. 11925408 and No. 12188101, the Ministry of Science and Technology of China (Grant No. 2018YFA0305700), the Chinese Academy of Sciences (Grant No. XDB33000000), the K. C. Wong Education Foundation (GJTD-2018-01), the Beijing Natural Science Foundation (Z180008), and the Beijing Municipal Science and Technology Commission (Z191100007219013).

## Notes

The authors declare no competing financial interest.

## Data availability

All the data and methods are present in the main text and the supplementary materials. Any other relevant data are available from the authors upon reasonable request.

## Additional information

**Supplementary information:** The online version contains supplementary material available at https://doi.org/..........